\documentclass[%
	amsmath,amssymb,
	aip,
    jcp,
    reprint,
    nolongbibliography
]{revtex4-1}
\usepackage{graphicx}

\newcommand{\dee}{\text{d}}
\newcommand{\avg}[1]{\langle {#1} \rangle}
\renewcommand{\vec}[1]{{\mathbf{#1}}}
\newcommand{\vr}{\vec r}
\newcommand{\vri}{\vec r_i}

\newcommand{\vex}{{v}}
\newcommand{\barvex}{{{\bar v}}}
\newcommand{\vexi}{{v_i}}
\newcommand{\barvexi}{{{\bar v}_i}}

\newcommand{\vrin}{{\vec r\in \vex}}
\newcommand{\vrPin}{{\vec r'\in \vex}}
\newcommand{\vrPPin}{{\vec r''\in \vex}}
\newcommand{\vrPPPin}{{\vec r'''\in \vex}}
\newcommand{\vrout}{{\vec r\in \barvex}}
\newcommand{\vrPout}{{\vec r'\in\barvex}}
\newcommand{\vrPPout}{{\vec r''\in\barvex}}
\newcommand{\vrPPPout}{{\vec r'''\in\barvex}}

\newcommand{\chiin}{\chi_\vex}
\newcommand{\chim}{\chi^{\text{(m)}}}

\newcommand{\Hnorm}{H_{\text{norm}}}
\newcommand{\Heff}{H_{\text{eff}}}

\renewcommand{\dee}{\text{d}}
\newcommand{\kB}{k_{\text{B}}}

\begin{document}

\title{An improved coarse-grained model of solvation and the hydrophobic effect}
\author{Patrick Varilly}
\author{Amish J. Patel}
\author{David Chandler}
\email{chandler@berkeley.edu}
\affiliation{Department of Chemistry, University of California, Berkeley, California 94720}
\date{\today}

\begin{abstract}
We present a coarse-grained lattice model of solvation
thermodynamics and the hydrophobic effect that implements the ideas of
Lum-Chandler-Weeks (LCW) theory [J.~Phys.~Chem.~B \textbf{103}, 4570
(1999)] and improves upon previous lattice models based on it.
Through comparison with molecular simulation, we show that our model
captures the length-scale and curvature dependence of solvation free
energies with near-quantitative accuracy and two to three orders
of magnitude less computational effort, and further, correctly
describes the large but rare solvent fluctuations that are involved in dewetting, vapor tube formation and
hydrophobic assembly.  Our model is intermediate in detail and
complexity between
implicit-solvent models and explicit-water simulations.
\end{abstract}

\maketitle

\section{Introduction}
\label{sec:intro}

This is a technical paper that addresses how the hydrophobic
effect may be understood quantitatively.  Despite its technical
nature, the physical ideas and final model we
formulate should be accessible and potentially interesting to a wide audience
of researchers who are confronted with the many manifestations of the
hydrophobic effect, and are in need of an effective quantitative tool for
treating them.

The hydrophobic effect is presumed to be an important driving force in biology and nanoscale
self-assembly.\cite{Kauzmann1959,Tanford1973,AlbertsEtAl2007}  Because
of its collective nature and its length-scale
dependence,\cite{Chandler2005} and because of its
nonlocal dependence on solute surface moeities,\cite{ChennamsettyEtAl2009,ChennamsettyEtAl2010,AcharyaVembanurJamadagniGarde2010} modeling
the hydrophobic effect remains a challenge.  To treat it
theoretically, one could track the explicit position of
possibly tens of thousands of water molecules around solutes of
interest,\cite{ShirtsPiteraSwopePande2003,ShirtsPande2005} but the computational cost of this approach limits its
applications.  Alternatively, at significantly
reduced cost, one could
replace explicit waters by an implicit solvent model, as is done in
the generalized Born and surface area (GBSA) approach.\cite{GBSA90,GBSA97}  In this paper, building on previous
efforts,\cite{LumChandlerWeeks1999,tenWoldeSunChandler2001,tenWoldeChandler2002} we
propose a coarse-grained model intermediate in detail between these
two extremes, one that retains most of the computational
advantage of implicit solvent models and overcomes two of their
significant conceptual flaws: their incorrect scaling behavior and
their neglect of rare but large solvent density fluctuations that play
pivotal roles in the dynamics of assembly.

Solvation free energies\cite{Chandler1987,RouxSimonson1999}
of solutes with sub-nanometer features, exactly the size prevalent in
biological regimes, do not in general\footnote{For reduced classes of solutes, such as
linear alkanes, surface-area scaling is nonetheless observed.  These
molecules are properly in the small length-scale regime, where
solvation free energy scales as volume.  However, for linear
molecules, surface area also scales as volume, leading to the
misleading scaling behavior.\cite{Chandler2005}  Moreover, the
resulting empirical surface tension is almost negligibly
small.  Typical values are in the $5$--$10\,$cal/mol/\AA${}^2
  \approx 1\,\kB T$/nm${}^2$
  range,\cite{ChenBrooksKhandogin2008,GBSA97} in contrast to the
  water-air surface tension of about~$17\,\kB T$/nm${}^2$ and
  the water-oil surface tension of about~$12\,\kB T$ nm${}^2$: see Ref.~\onlinecite{Tanford1979}.}\llap{\phantom{\cite{Tanford1979}}} scale as surface area.\cite{LumChandlerWeeks1999,HGC01,LevyEtAl2003}  By construction, models that assume such scaling
significantly underestimate the driving force for hydrophobic
assembly.\cite{Chandler2005,ChenBrooks2007}  Our model, on the
other hand, captures the correct scaling behavior with solute size for generic solute geometries.

Since hydrophobicity is a solvent property as much as it is a solute
property, it is important to consider the solvent on length scales dictated by the solute(s).
Numerous studies of hydrophobicity\cite{WallqvistBerne1995,LumLuzar1997,BolhuisChandler2000,tenWoldeChandler2002,AnishkinSukharev2004,Chandler2005,MelittinNature2005,MillerVandenEijndenChandler2007,AthawaleEtAl2007,RasaiahGardeHummer2008,JamadagniEtAl2009,BerneWeeksZhou2009,PatelVarillyChandler2010}
have shown that rare solvent motions and dewetting transitions in
confining environments play a critical
role in solute assembly and function.  Our model adequately captures these rare and
important fluctuations.  To demonstrate this, we consider the water number
distribution~$P_V(N)$ in a probe volume~$V$.  Hummer et al.~introduced
the idea of characterizing this distribution in the context of
solvation theory,\cite{HummerEtAl1996} and the utility of this
function has been demonstrated subsequently.\cite{[{See, for
  instance, }][{ and Refs.~\onlinecite{GardeKhareHummer2000}, \onlinecite{PatelVarillyChandler2010}~and~\onlinecite{XuMolinero2010}.}]HuangChandler2000b}\llap{\phantom{\cite{GardeKhareHummer2000,XuMolinero2010}}}

The GBSA model,\cite{GBSA90,GBSA97} widely used in biological
settings, captures the effect of electrostatics with reasonable
accuracy, but its treatment of the hydrophobic effect is less
adequate,\cite{LevyEtAl2003,Zhou2003,DaidoneEtAl2007,ChenBrooksKhandogin2008}
for the reasons discussed above.  Interesting examples of solutes for
which hydrophobicity is essential, and for which GBSA is unsuitable,
include large classes of proteins, such as those involved in
transmembrane protein recognition and insertion,\cite{SRPNature2010}
and versatile chaperones.\cite{Hsp90Structure}  It is these kinds of
solutes for which our approach may eventually prove most useful.

The ideas behind our model are those of Lum-Chandler-Weeks
(LCW)\cite{LumChandlerWeeks1999} theory.  Ten Wolde, Sun and
Chandler\cite{tenWoldeSunChandler2001,tenWoldeChandler2002}
generalized this theory by casting it in terms of a Hamiltonian for a
lattice field theory.  The motivation for that development was to
facilitate treatments of large length scale dynamics.  The
motivation of the current work is similar, though in this paper we
confine our attention to time-independent properties.  The main
contribution of this paper is to improve upon
these previous attempts, and to introduce concrete implementations of the
underlying theory that illustrate the improvements, which are significant.

The paper is organized as follows.  In Section~\ref{sec:model}, we
sketch the physical ideas behind our model and present
their implementation in a tractable format.  The derivation and approximations made therein are
left to the Appendix.  In Section~\ref{sec:results}, we consider the accuracy of our model by computing the solvation free energies of
solutes and $P_V(N)$ distributions for various geometries, with and without adhesive
solute-solvent interactions.  Finally, in Section~\ref{sec:discussion} we
conclude with a discussion of the merits and limitations of the
present implementation of the model.

\section{Model}
\label{sec:model}

\subsection{Density fields and Hamiltonian}

In this section we first consider general features of a liquid
solvent, specializing to water only later.  We focus on the solvent
density, $\rho(\vr)$.  For water in particular,
$\rho(\vr)$ refers to the instantaneous
positions of water oxygen atoms.  Effects of other variables such as molecular orientations
appear implicitly in terms of parameters.  We decompose density into large and
small length-scale contributions $\rho_\ell
n(\vr)$~and~$\delta\rho(\vr)$, respectively,
\begin{equation}
\rho(\vr) = \rho_\ell n(\vr) + \delta\rho(\vr).
\label{eqn:partsOfRho}
\end{equation}
Here, $\rho_\ell$ is the bulk liquid density, and $n(\vr)$ is an Ising-like field that is $1$ in
regions that are locally liquid-like and $0$ in regions that are
locally vapor-like.  The field takes on intermediate values
  only around interfacial regions.
This large length-scale field describes extended liquid-vapor interfaces, while
the small length-scale field describes more rapidly-varying
density fluctuations.  This separation implies some form of space-time
coarse-graining to define~$n(\vr)$, a coarse-graining which is most reasonable for dense
fluids far from their critical points.  The key development of LCW was to describe how
to (a) perform this decomposition, and (b) couple the two
separate fields.

Building on the work of ten Wolde, Sun and
Chandler,\cite{tenWoldeSunChandler2001} we construct a Hamiltonian for
the solvent density that captures the dominant physics.  We have
\begin{multline}
H[n(\vr),\delta\rho(\vr)] = H_{\text{large}}[n(\vr)]\\
+ H_{\text{small}}[\delta\rho(\vr); n(\vr)]
+ H_{\text{int}}[n(\vr),\delta\rho(\vr)]. \label{eqn:HMicro}
\end{multline}
The term $H_{\text{large}}[n(\vr)]$ captures the physics of interface
formation in~$n(\vr)$.  For the term~$H_{\text{small}}[\delta\rho(\vr);n(\vr)]$, we exploit
the observation that small length-scale density fluctuations in homogeneous liquids obey Gaussian
statistics.\cite{Chandler1993,HummerEtAl1996,CrooksChandler1997}
Thus, for a given configuration of~$n(\vr)$, we assume
that~$\delta\rho(\vr)$ has Gaussian statistics with variance
$$
\chi(\vr, \vr'; [n(\vr)]) = \avg{\delta\rho(\vr) \delta\rho(\vr')}_{n(\vr)}.
$$
Here, the right-hand side denotes the thermal average of
$\delta\rho(\vr)\delta\rho(\vr')$ under the constraint of fixed~$n(\vr)$.
The term $H_{\text{small}}[\delta\rho(\vr);n(\vr)]$ is then a Gaussian with this variance,
namely
\begin{multline}
H_{\text{small}}[\delta\rho(\vr); n(\vr)] = \\
\frac{\kB T}{2} \int_\vr
\int_{\vr'} \delta\rho(\vr) \chi^{-1}(\vr,\vr';[n(\vr)])
\delta\rho(\vr'),
\label{eqn:Hsmall}
\end{multline}
where $\kB T$ is temperature, $T$, times Boltzmann's constant.
For conciseness, we use an abbreviated integration notation,
where the integration variable is denoted with a subscript to the
integral sign, and the integration domain is all of space unless otherwise stated.
We approximate the variance with
\begin{equation}
\chi(\vr,\vr';[n(\vr)]) \approx 
\begin{cases}
\chi_0(\vr - \vr'),&n(\vr) = n(\vr') = 1;\\
0,&\text{otherwise},
\end{cases}
\label{eqn:chi}
\end{equation}
where $\chi_0(\vr)$ can be written in terms of the radial distribution
function $g(r)$ as
\begin{equation}
\chi_0(\vr) = \rho_\ell \delta(\vr) + \rho_\ell^2 [g(|\vr|) - 1].\label{eqn:chi0}
\end{equation}
For the uses we make of the approximation in Equation~\eqref{eqn:chi}, corrections have
quantitative but not qualitative effects, as discussed in
the Appendix and also Ref.~\onlinecite{[][{, Figure (3.6),
    page~61.}]Lum1998}.  Finally, $H_{\text{int}}[n(\vr),\delta\rho(\vr)]$ is an effective
coupling between $n(\vr)$~and~$\delta\rho(\vr)$ due to unbalanced
attractive forces in the solvent, whose details are given in the Appendix.

In the absence of large solutes, fluctuations in $n(\vr)$ are unlikely.  The only
fluctuations of significance in that case are those described by
$\delta\rho(\vr)$.  In the presence of large solutes, however, $n(\vr)$ will often differ
significantly from its bulk mean value.  In that case the statistics
of $\delta\rho(\vr)$ is modified and the coupling~$H_{\text{int}}[n(\vr),\delta\rho(\vr)]$ between
$n(\vr)$~and~$\delta\rho(\vr)$ becomes significant.  When $\delta\rho(\vr)$ is integrated
out, a renormalized Hamiltonian for $n(\vr)$ results.

LCW theory~\cite{LumChandlerWeeks1999} is a mean-field theory for the
average large length-scale field, $\langle n(\vr)\rangle$, so it ignores
the effects of large-scale fluctuations in $n(\vr)$.  Subsequent
lattice implementations of LCW
theory~\cite{tenWoldeSunChandler2001,tenWoldeChandler2002,WillardChandler2008}
have incorporated fluctuations in the simplest possible manner.  The
present model refines these previous attempts to achieve
near-quantitative accuracy for solvation free-energies and correct
behavior of fluctuations in~$n(\vr)$.  Most importantly, we improve the calculation of the interfacial energies
due to~$n(\vr)$.

\begin{figure}
\includegraphics{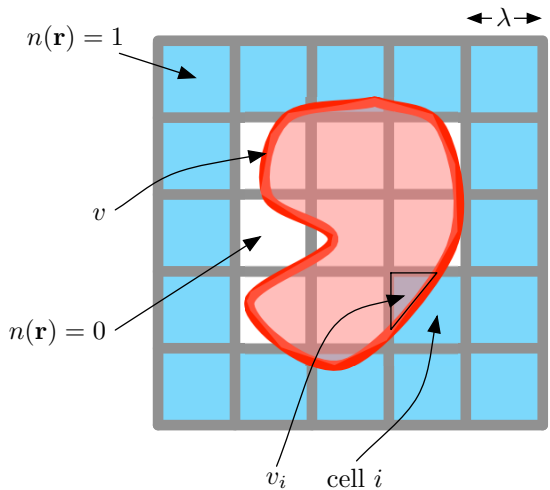}
\caption{\label{fig:cartoon}Schematic showing the solute and the large
  length-scale density field on a grid}
\end{figure}

To write down the renormalized Hamiltonian, we begin by describing $n(\vr)$
with reference to a cubic grid of spacing~$\lambda$, depicted in
Figure~\ref{fig:cartoon}, and we denote its value at the center of cell~$i$ by
$n_i$.  Then, $n(\vr)$ is given by
\begin{equation}
n(\vr) = \sum_i n_i \Psi(\vr - \vri),
\label{eqn:nofr}
\end{equation}
where $\vri$ is the center of cell~$i$ and $n_i$ is $1$~or~$0$, and
the sum is over all cells~$i$.  The function
$\Psi(\vr)$ is maximal with value~$1$ at $\vr=\vec 0$; it is cubic
symmetric about the origin; and it is zero when the magnitude of any of the
Cartesian components of~$\vr$ is greater than~$\lambda$.  The value of $\lambda$, about which we will say more
later, should be roughly the size of the bulk
correlation length of the liquid solvent.  The typical size of interfacial energies between
cells on this grid is~$\gamma\lambda^2$, where $\gamma$ is the
liquid-vapor surface tension of the solvent.  The dissolved solute excludes
solvent density from a volume~$\vex$, and we define $\barvex$ to be
its complement, so that the total volume of the system is the union of
$\vex$~and~$\barvex$.  The excluded volume can be of any shape, and
it can be composed of disconnected parts.  In
Ref.~\onlinecite{Chandler1993}, regions within $\vex$ are called
``in'', and regions within $\barvex$ are called ``out''.  For any
volume~$V$, we have the projector~$b_V(\vr)$,
\begin{equation}
b_V(\vr) = \begin{cases}
1,&\vr \in V,\\
0,&\text{otherwise},
\end{cases}
\end{equation}
so that $b_V(\vr) + b_{\bar V}(\vr) = 1$.
We denote the overlap of $\vex$~or~$\barvex$ with cell~$i$ by
$\vexi$~and~$\barvexi$, respectively.

In the absence
of a solute, the Gaussian nature of $\delta\rho(\vr)$ results in
solvent number fluctuation correlations.  The correlations between the
portions of cells $i$~and~$j$ that overlap with two volumes
$V$~and~$V'$, respectively, form the elements of a matrix
\begin{equation}
\chi_{ij}(V,V') = \int_{\vr \in i} \int_{\vr' \in j}
b_V(\vr) \chi_0(\vr,\vr') b_{V'}(\vr').\label{eqn:Aij}
\end{equation}
Here, the domain of the $\vr$~and~$\vr'$ integrals are restricted to the volume of
cells~$i$~and~$j$, as indicated.
A way to estimate these elements is outlined in the Appendix.  The resulting matrix is used to calculate entropic effects due to solvent exclusion
and the linear response of solvent density to external fields.

Part of the renormalized Hamiltonian is the free energy~$H_{\text{large}}[n(\vr)]$ of the field~$n(\vr)$ in the absence of external solutes.  We estimate this contribution using a Landau-Ginzburg Hamiltonian
\begin{equation}
H_{\text{large}}[n(\vr)] = \int_{\vr} \Bigl[ w(n(\vr),\mu) + \frac{m}{2} |\vec\nabla n(\vr)|^2 \Bigr],\label{eqn:HLContinuum}
\end{equation}
where $w(\rho/\rho_\ell,\mu)$ is the grand free energy density for the
liquid solvent at a given density, $\rho$, and chemical potential,
$\mu$, relative to that of the gas.  The parameter~$m$ reflects
surface tension and intrinsic interfacial width.  At liquid-gas phase
coexistence, $\mu = 0$, the value of the integral is conveniently
expressed as the sum $\gamma\lambda^2\sum_i h_i$ with the local integrals
\begin{multline}
h_i = \frac{1}{\gamma\lambda^2}
\int_{x_i}^{x_i+\lambda}\dee x\, 
\int_{y_i}^{y_i+\lambda}\dee y\, 
\int_{z_i}^{z_i+\lambda}\dee z\\
\Bigl[ w(n(x,y,z),0) + \frac{m}{2} |\vec\nabla n(x,y,z)|^2 \Bigr].
\label{eqn:hi}
\end{multline}
The quantity~$h_i$ depends only on the values of $n_j$ for cells~$j$ that share one of the corners of cell~$i$.  There are only $14$ distinct possible values of $h_i$, which can be precalculated numerically for a given free energy density $w(n,0)$ and cell size~$\lambda$, as detailed in the Appendix.  In previous
modeling, the simpler Ising model
estimate~$\gamma\lambda^2 \sum_{\langle ij\rangle} (n_i - n_j)^2$ has
been used.
For reasons discussed in the Appendix, this simpler estimate
proves less accurate than the one used here.

\begin{subequations}
 With the above notation, we now write the Hamiltonian for our model,
which constitutes the main result of the paper,
\begin{multline}
\Heff[\{n_i\}] = \gamma\lambda^2 \sum_i h_i  - \mu
\rho_\ell\lambda^3 \sum_i n_i\\
+ K \sum_i \phi_i (-\rho_\ell n_i \vexi)\\
+ \kB T [ \avg{N}_v^2 / 2 \sigma_v + C/2 ],\label{eqn:Heff}
\end{multline}
where
\begin{align}
\phi_i &= 2a\rho_\ell \Bigl[1 - \frac12 n_i - \frac1{12}
\sideset{}{^\prime}\sum_{j\,\text{(nn$i$)}} n_j\Bigr],\label{eqn:discretePhii}\\
\avg{N}_v &= \rho_\ell \sum_i n_i v_i,\label{eqn:discreteAvgNv}\\
\sigma_\vex &= \sum_{i,j} n_i \chi_{ij}(\vex,\vex)
n_j,\label{eqn:discreteSigmaV}\\
C &= \begin{cases}
\ln(2\pi\sigma_v),&\avg{N}_v > 1,\\
\max[ \ln(2\pi\sigma_v), \avg{N}_v ],&\text{otherwise}.
\end{cases}\label{eqn:C}
\end{align}
The field $\phi_i$, with strength governed by the positive
parameter~$a$, is what Weeks has termed an ``unbalancing
potential''.\cite{Weeks2002,WeeksSelingerBroughton1995,WeeksVollmayrKatsov1997,WeeksKatsovVollmayr1998,KatsovWeeks2001}
The primed sum over $j\text{(nn$i$)}$ is a sum over the six cells~$j$ that are
nearest neighbors to cell~$i$.  The final expression for $\phi_i$
shown above, with renormalization constant~$K$, is the result of an
accurate and computationally convenient approximation, which is described in detail in the Appendix.
\label{eqn:model}
\end{subequations}

The terms on the right-hand-side of Equation~\eqref{eqn:Heff} respectively approximate: the free-energy cost of
establishing interfaces in $n(\vr)$, the pressure-induced bias towards
the liquid phase, the effective coupling between
$n(\vr)$~and~$\delta\rho(\vr)$ induced by the presence of a solute,
and the entropic cost of excluding solvent density from the portions
of~$\vex$ where $n(\vr) = 1$.  As the total number of waters to be
excluded, $\avg{N}_v$, approaches zero, the statistics of solvent number
fluctuations in~$v$ changes from Gaussian to Poisson, so that its
variance, $\sigma_v$, also approaches zero.
Equation~\eqref{eqn:C} captures this change continuously and prevents $\Heff[\{n_i\}]$
from becoming infinitely negative in this limit.

\subsection{Incorporating Solute-solvent attractions}

The above Hamiltonian pertains to the simplest case, where the solute interacts with the
solvent only by hard-core repulsion.  Realistic solutes additionally
have attractive interactions with the solvent that can be modeled as
an external potential~$u(\vr)$ that couples to~$\rho(\vr)$.  Such a
potential induces an additional term in our Hamiltonian, which we denote
by~$H_u[\{n_i\}]$.  To describe this term, we define a discretized analogue~$u_i$ of $u(\vr)$,
\begin{equation}
u_i = \frac{1}{\barvexi}\int_{\vr\in i} b_{\barvex}(\vr) u(\vr).
\label{eqn:ui}
\end{equation}
Notice that $u_i$ is independent of $u(\vr)$ for values of $\vr$
inside the solute.  The apparent divergence, where $\vex$ completely
overlaps cell~$i$, has no effect in the final expression.  In
particular,
\begin{multline}
H_u[\{n_i\}] =\\
\sum_i u_i n_i \Bigl[ \rho_\ell \bar \vexi - \sum_j
\chi_{ij}(\barvex, \vex) n_j \avg{N}_v / \sigma_v \\
- \sum_j \chi_{ij}(\barvex, \barvex) n_j \beta (u_j + \phi_j) \Bigr]\\
+ \frac{\kB T}2 \sum_{i,j} \beta u_i n_i \chi_{ij}(\barvex,\barvex)
n_j \beta u_j,
\label{eqn:HuDiscrete}
\end{multline}
where $\beta$ is the reciprocal of $\kB T$.
The first part is the mean-field contribution~$\int_\vr u(\vr)
\langle\rho(\vr)\rangle$, while the last term is the entropic cost of the external potential modifying the average solvent density in the vicinity of the solute.

\subsection{Parameters of the Hamiltonian}

We now specialize our model to water at ambient conditions, $T=298\,$K
and $1\,$atm pressure, $p$.  Further, we comment upon what changes are
required for applications at different states of water.

The cell size length~$\lambda$ should be no smaller than the intrinsic
width of the liquid-vapor interface.  Based upon the interfacial
profile of the SPC/E model,\cite{BerendsenGrigeraStraatsma1987,HGC01}
we therefore pick~$\lambda = 4\,\text{\AA}$.
This is the minimal scale over which the time-averaged solvent density can transition from liquid-like to vapor-like values.  Following Ref.~\onlinecite{HuangChandler2002}, we use the free-energy density
\begin{equation}
w(n,\mu) = \frac{2m}{d^2} (n - 1)^2 n^2 - \mu \rho_\ell n,
\label{eqn:omega}
\end{equation}
with $d = 1.27\,$\AA\ because this choice reproduces the sigmoidal
density profile of water-vapor interfaces at coexistence.  The resulting values of $h_i$ are tabulated in the Appendix.  The bulk liquid density~$\rho_\ell$ is the
experimental value,\cite{NISTWater} whereby a liquid cell contains
$\rho_\ell \lambda^3 \approx 2.13$ waters on average.  The value of~$m$ is chosen such that the interfacial
energy of vapor spheres of radius~$R$ tends to $4\pi \gamma R^2$ for
large~$R$.  At ambient conditions,
the experimental value for the surface tension\cite{NISTWater} yields $\gamma\lambda^2 \approx 2.80\,\kB T$.
Finally, the relative chemical potential is given by
$\mu \approx (p - p_{\text{vap}})/\rho_\ell$, where $p_{\text{vap}}$
is the vapor pressure at~$298\,$K.  This relationship gives $\mu
\approx 7.16\times10^{-4}\,\kB T$, which is quite small, reflecting
that water at ambient conditions is nearly at coexistence with its
vapor. 

The matrix elements $\chi_{ij}(V,V')$ are computed from the radial
distribution function, $g(r)$, and we derive this function from Narten
and Levy's tabulated data.\cite{NartenLevy1971}  It is a convenient
data set because it covers a broad range of temperatures for the liquid at and near $p =
1\,$atm.  At one temperature, 25$^\circ$C, we have checked that a
different estimate of the radial distribution function, that of the
SPC/E model, yields similar matrix elements, and the resulting
solvation properties are essentially identical to those obtained when
the $\chi_{ij}(V,V')$'s are computed from the Narten-Levy data at the
same temperature.

The only parameters that we estimate through fitting are the
strength~$a$ of the unbalancing potential and the renormalization
constant~$K$.  In the absence of solute-solvent attractions, only the
product of $a$~and~$K$ is relevant.  Values of $a$~and~$K$ with $Ka\rho_\ell = 2.1\,\kB T$
allow us to match the solvation free
energies of hard spheres in SPC/E water (see below).  By comparing the
average value of the computationally convenient approximate expression involving $\phi_i$ in
Equation~\eqref{eqn:Heff} with that of its complete and unrenormalized
counterpart, as is done in the Appendix, we find that $K$ is about
$1/2$, so that $a\rho_\ell \approx 4.2\,\kB T$.
This value for $a$ is close to the original LCW
estimate,\cite{LumChandlerWeeks1999} arrived at from a
different criterion.

These values are applicable at ambient conditions.  As temperature and
pressure vary, only $\gamma$, $\mu$ and $g(r)$ vary appreciably, while
$K$ varies slightly.  In
particular, surface tension decreases roughly linearly with
temperature\cite{NISTWater} (with $\dee\gamma/\dee T \approx
-0.15\,$mJ/m$^2\cdot$K which is $-5.8\times 10^{-3}\,\kB
T/\lambda^2\cdot$K at $T = 298\,$K).  As noted above, $\mu$ increases
roughly linearly with pressure.  The pair correlation function $g(r)$
loses some structure for temperatures above~$50\,{}^\circ$C.
The terms that are modeled by the renormalization constant~$K$ reflect the
degree to which solvent density layers next to a solute.  Since this layering
reflects the structure of~$g(r)$, we expect $K$ to be
slightly state-dependent, with its value increasing with temperature.

Conversely, liquid water has a nearly constant density and bulk
correlation length at the temperatures and pressures where our model
would be useful, so $\rho_\ell$~and~$\lambda$ can be taken as constant
as well.  Theoretical estimates for $a$ in simple liquids (Eq.~4 in
Ref~\onlinecite{Weeks2002}) are state-independent, so we expect that
in water, $a$ will be nearly state-independent as
well.\cite{HuangChandler2000a}

\section{Applications and Results}
\label{sec:results}

\subsection{Solvation Free Energies}
\label{sec:results:DeltaG}

To test our model's ability to capture the length-scale dependence of
solvation, and to parametrize the strength of the unbalancing potential, we have calculated the solvation free energy of hard spheres of different radii.
Whether within our model or using explicit water simulations, we calculate the solvation free energy of a solute following the
guidelines of
Ref.~\onlinecite{PohorilleJarzynskiChipot2010}.  Briefly, we first
define a series of~$M+1$ solutes $S_0$~through~$S_M$ that slowly interpolate
from an empty system ($S_0$) to the final solute of interest ($S_M$).
We then sequentially calculate the free energy difference between
solute~$m$ and solute~$m+1$ using the Bennett acceptance ratio
estimator\cite{Bennett1976} (BAR), and, where necessary, the linear
interpolation stratification procedure of
Ref.~\onlinecite{PohorilleJarzynskiChipot2010}.  Error estimates are calculated using
BAR, and are generally smaller than $0.5$\%.

\begin{figure}
\includegraphics{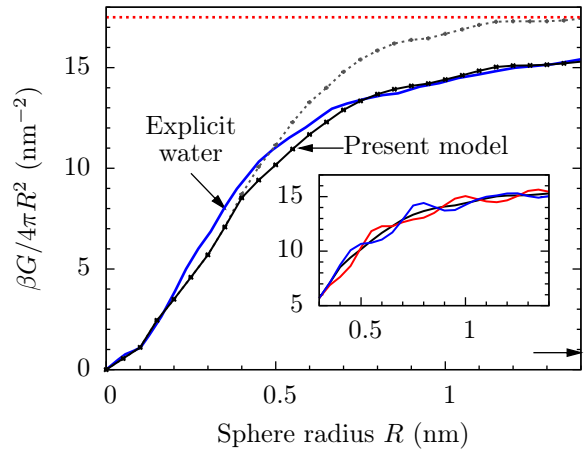}
\caption{Solvation free energies~$G$ of hard spheres of increasing
  radii, as calculated from explicit SPC/E water
  simulations\cite{HGC01} (solid blue), from the coarse-grained
  model (solid black), and from the most common GBSA
  variant\cite{ChenBrooksIII2008} (arrow at bottom
  right).  When the coarse-grained model has no
  unbalancing potential ($a=0$, dashed gray), the intermediate-size
  regime is only qualitatively reproduced.  For large spheres,
  the ratio of~$G$ to surface area tends to the liquid-vapor surface
  tension~$\gamma$ (horizontal red dots).  Inset: Illustration of
  lattice artifacts.  The spheres are centered at
  different offsets from the lattice: a generic position
  $(0.98\,\text{\AA},0.79\,\text{\AA},1.89\,\text{\AA})$ that breaks
all rotational and mirror symmetries (black), a lattice cell corner
(blue) and a lattice cell center (red).  All three curves are identical for $R \leq 0.35\,$nm.\label{fig:solvation_spheres}}
\end{figure}

Our model (Equation~\eqref{eqn:Heff}) involves only simple arithmetic,
so free energies can be calculated with little computational effort.
For example, calculating the solvation free energy of hard spheres of
up to $14\,$\AA\ in radius in increments of $0.5\,$\AA\ 
(Figure~\ref{fig:solvation_spheres}) takes about $1$ hour on a single
$2\,$GHz machine with a code that has not been fully optimized, whereas a similar calculation in explicit SPC/E
waters with GROMACS\cite{HessEtAl2008} would take around
$600$ hours on the same machine to obtain a similar statistical accuracy.
%
%

Hard-sphere solvation free energies scale as solute volume for small
spheres, and as surface area for large spheres, with a smooth
crossover at intermediate
sizes.\cite{Chandler2005} Figure~\ref{fig:solvation_spheres}
illustrates this behavior and compares the results of our model to
previous simulation results using SPC/E water.\cite{HGC01} As the model manifestly
reproduces the small- and large-length scale limits, the most
significant feature illustrated in Figure~\ref{fig:solvation_spheres}
is the \emph{gradual} crossover from volume to surface area scaling.
Ignoring the unbalancing potential leads to a qualitatively correct
scaling behavior.  However, adjustment of the single parameter~$a$,
which determines the strength of the unbalancing potential, yields a near-exact agreement between our model and the
SPC/E results for all sphere sizes.  In all subsequent results, the
parameter~$a$ is fixed at this value.

The model
results have small lattice artifacts---results that depend upon the position
of the solute relative to that of the coarse-grained lattice---as
shown in the inset of Figure~\ref{fig:solvation_spheres}.  When
studying stationary solutes, lattice artifacts may be mitigated by
performing multiple calculations, differing only by small
displacements of the solutes, and then averaging the results.  When
studying dynamical phenomena, lattice
artifacts tend to pin solutes into alignment with the coarse-grained lattice.  For arbitrary
molecular solutes, we expect that pinning forces acting on
one portion of the molecule will generically oppose pinning forces on
other parts of the molecule, so that the total pinning forces will
largely cancel out.  However, when treating many identical molecules,
lattice artifacts can add constructively, and additional steps are
needed to mitigate them.\cite{SztrumVartashRabani2010}

\begin{figure}
\includegraphics{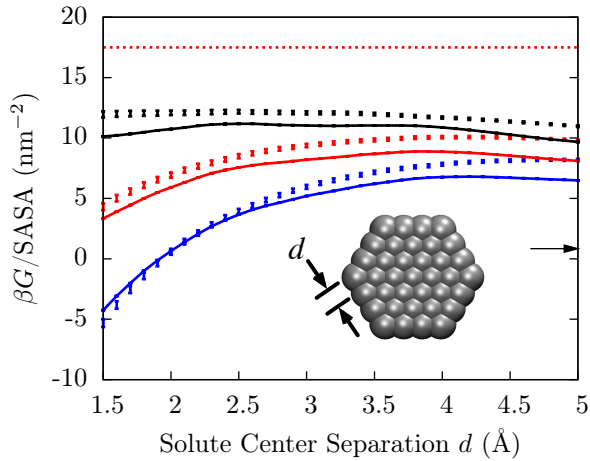}
\caption{\label{fig:solvation_hexplates}Solvation
  free energies~$G$ of hexagonal plates, as a function of plate size, as calculated by the coarse-grained
  model (solid lines), by explicit SPC/E water simulations (points),
  and by the most common GBSA variant (arrow on right).
  Three values of the attractive interaction strength~$\eta$ are
  shown: $0.0$ (black), $0.5$ (red) and $1.0$ (blue).  Solvent-accessible surface
  areas (SASAs) were calculated using
  VMD,\cite{[][{. VMD was developed
  by the Theoretical and Computational Biophysics Group in the Beckman
  Institute for Advanced Science and Technology at the University of
  Illinois at Urbana-Champaign.}]HumphreyDalkeSchulten1996} with a particle radius
  of~$1.97\,$\AA, a solvent radius of~$1.4\,$\AA, and
  1,000,000~samples per atom.  The bulk liquid-vapor surface tension
  of water (horizontal red dots) is shown.  Inset: Detail of the
  hexagonal plate.  The solvent-excluded volume of each oily
  site is a sphere of radius $R_0 = 3.37\,$\AA.}
\end{figure}

Since the unbalancing potential is explicitly parametrized with the
solvation free energy of hard spheres, it is useful to evaluate the
accuracy of the results in other geometries.  To this effect we
computed the solvation free energies of a family of hexagonal plates, consisting of
$37$~methane-like oily sites arranged into three concentric rings.  We control
the size of these plates, depicted in Figure~\ref{fig:solvation_hexplates}, by
varying the distance~$d$ between neighboring oily sites.  For our
calculations with explicit SPC/E water, the sites are uncharged and
interact with the solvent molecules via a standard\footnote{The
  parameters of the solute-solute Lennard-Jones potential are those of
  Ref.~\onlinecite{JorgensenMaduraSwenson1984}: $\sigma=3.905\,$\AA\
  and $\epsilon=0.118\,$kcal/mol.  Lorentz-Berthelot mixing rules were
  used to obtain the water-solute interaction
  parameters.}\llap{\phantom{\cite{JorgensenMaduraSwenson1984}}}
water-methane Lennard-Jones potential.  To study the role of
attractive interactions, we split this Lennard-Jones potential using
the Weeks-Chandler-Andersen (WCA) prescription\cite{WCA} into a
repulsive part~$u_0(r)$ and an attractive part~$\Delta u(r)$.  The
magnitude of the attractive tail can be varied systematically with a
scaling parameter~$\eta$, such that
\begin{equation}
u_\eta(r) = u_0(r) + \eta \Delta u(r).
\label{eqn:defn_lambda}
\end{equation}
For the ideal hydrophobic plate, we set $\eta$ to zero.

In the coarse-grained model, the repulsive core of the solute is represented
as an excluded volume.  To construct it, we replace each solute
particle by a thermally-equivalent hard sphere, whose radius~$R_0$ is
estimated according to
$$
R_0 = \int_0^\infty \dee r\, [1 - \exp(-\beta u_0(r))],
$$
which is a first approximation to the WCA value of this
radius,\cite{[][{, Section 5.3}]HansenMcDonald2006,VerletWeis1972} and
is essentially the radius at which $u_0(r)$ is~$\kB T$.
The excluded volume is then the union of the hard-sphere volumes of
each solute site.

Figure~\ref{fig:solvation_hexplates} compares the solvation free
energies for this family of solute plates computed from our atomistic
simulations with those computed from the coarse-grained model with the
unbalancing parameter~$a$ determined above for solvated hard spheres.
Now, with this different geometry, the coarse-grained model continues
to perform well.  The discrepancies are primarily due to
the small underestimation, shown in Figure~\ref{fig:solvation_spheres}, of the solvation free energy of small spheres.
Figure~\ref{fig:solvation_hexplates} also compares the solvation free
energies of plates with increasing attractions to
the corresponding results from explicit-water simulations.  Aside from the small artifacts
already present in the ideal solute case, the contribution
of the attractions to solvation free energies calculated with the coarse grained model is nearly quantitative.

\subsection{Fluctuations}
\label{sec:results:PvN}

A more detailed probe of solvent behavior than solvation free energies
is the probability~$P_V(N)$ of finding $N$~waters in a given
volume~$V$.  The solvation free energy~$G$ of an ideal,
volume-excluding hydrophobe is
simply\cite{HummerEtAl1996}~$\beta G = -\ln P_V(0)$, and we can glean information about hydrophobicity and dewetting from the behavior at non-zero~$N$.

In the present model, we estimate $P_V(N)$ by a two-step procedure.
For any given solvent configuration~$\{n_i\}$, the small length-scale
fluctuations of~$\delta\rho(\vr)$ give rise to a Gaussian distribution
in the numbers of waters, so that
\begin{equation}
P_V(N | \{n_i\}) \propto \exp\bigl[ - (N - \avg{N}_V)^2 / 2 \sigma_V \bigr],
\label{eqn:PvNni}
\end{equation}
where,
\begin{multline}
\avg{N}_V = \sum_i n_i \Bigl[ \rho_\ell V_i - \sum_j \chi_{ij}(V,\vex) n_j \avg{N}_v/\sigma_v\\
-\sum_j \chi_{ij}(V,\barvex) n_j \beta (u_j + \phi_j) \Bigr],
\label{eqn:avgNni}
\end{multline}
and,
\begin{equation}
\sigma_V = \sum_{ij} n_i \chi_{ij}(V,V) n_j.
\end{equation}
Here, $V_i$ is the overlap of the probe volume with cell~$i$.  Notice the use of the probe volume $V$ in the $\chi_{ij}$ matrices.  Formally, we then thermally average the above result over all possible solvent configurations to obtain
$$
P_V(N) \propto \sum_{\{n_i\}} P_V(N | \{n_i\}) \exp(-\beta\Heff[\{n_i\}]).
$$
In practice, we estimate this sum by sampling a lattice
variable~$n$ that closely correlates with~$N$, given by
$$
n = \sum_{i\in V} n_i.
$$
We divide the range of possible values of $n$ into small
overlapping windows,
and sample relevant configurations at every value of~$n$ using
Wang-Landau sampling\cite{WangLandau2001} along~$n$, together with
replica exchange,\cite{EarlDeem2005} to obtain good sampling and avoid
kinetic traps.  We then used the multistate Bennet
acceptance ratio estimator\cite{MBAR} (MBAR) to
reconstruct from these runs the probability distribution~$P(n)$.
During the umbrella sampling runs, lattice configurations with equal~$n$ are observed in proportion to their Boltzmann weight.  Using the notation $\{n_i\} \in n$ to denote all observed lattice gas configurations with a particular value of~$n$, we finally obtain
$$
P_V(N) = \sum_{n} P(n) \sum_{\{n_i\} \in n} P_V(N | \{n_i\}).
$$
To estimate the statistical errors in our procedure, we calculate~$P_V(N)$ in five independent Monte Carlo runs, and estimate the standard error in the mean of $\ln P_V(N)$.

For comparison, we also calculate these distributions in SPC/E water using LAMMPS\cite{LAMMPS}
as described previously,\cite{PatelVarillyChandler2010} paying careful
attention to good sampling around free energy
barriers.  Errors were estimated
with MBAR.

\begin{figure}
\includegraphics{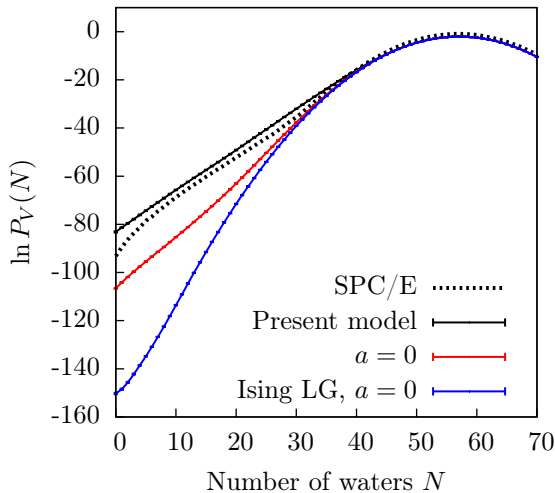}
\caption{Water number distribution in a
  $12\times12\times12\,$\AA${}^3$, as obtained using explicit SPC/E water simulations,\cite{PatelVarillyChandler2010}
  the present model, the present model without the unbalancing
  potential ($a=0$), and a model with an Ising Lattice Gas and no unbalancing potential.\label{fig:PvNSimpleVols}}
\end{figure}

In the absence of a solute, $P_V(N)$ is sensitive only to the
interfacial energetics of the lattice gas.
Figure~\ref{fig:PvNSimpleVols} compares the $P_V(N)$ curve obtained
using the present model for a $12\times12\times12\,$\AA${}^3$ volume
with results that we have previously obtained from simulation of SPC/E
water,\cite{PatelVarillyChandler2010} and with (a) a version of the
coarse-grained model that lacks an unbalacing potential ($a$~is set to
zero), and (b) a version that additionally uses the naive Ising
lattice gas for estimating interfacial energetics in~$n(\vr)$.  Our
present model captures the observed deviations from Gaussian behavior
better than these simpler models, which reflects its higher accuracy
in estimating interfacial energetics and microscopic curvature effects.

We have also previously examined how hydrophobic solutes affect water
number fluctuations in nearby probe volumes.\cite{PatelVarillyChandler2010}  To evaluate the performance of our
model in that scenario, we use the model hydrophobic plate solute
described in Ref.~\onlinecite{PatelVarillyChandler2010}.  The plate is
made up of oily particles with the same number density as water, whose
centers lie inside a $24\times24\times3\,$\AA${}^3$
volume\cite{SuppRef}.  Taking into account the van der Waals radii of the oily
particles, the plate has approximate dimensions
$28\times28\times7\,$\AA${}^3$.  We model this solute in the same way
as the hexagonal plates described above.  As before, we explore the
role of attractive interactions by varying the attraction strength parameter~$\eta$.

\begin{figure}
\includegraphics{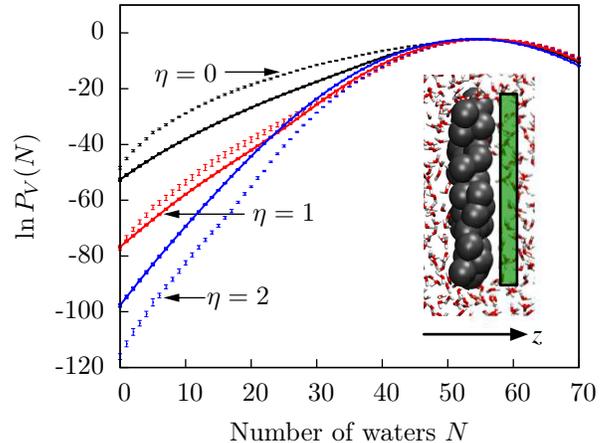}
\caption{Water number distributions in a probe volume of size
  $24\times24\times3\,$\AA${}^3$ immediately adjacent to a model plate
  solute (inset) of varying attractive strength~$\eta$, in the coarse-grained
  model (solid lines) and in explicit SPC/E water (points).  Defining
  $z=0$ to be the plane passing through the plate center, points in the
probe volume (green) satisfy $5\,\text{\AA} < z < 8\,\text{\AA}$, so that
  a water molecule touching the plate is located at the edge of the
  probe volume.\cite{PatelVarillyChandler2010} \label{fig:PvNNextToPlate}}
\end{figure}

 Figure~\ref{fig:PvNNextToPlate} shows the water number distribution in
a $24\times24\times3\,$\AA${}^3$ probe volume adjacent to the
plate.  With no solute-solvent attractive interactions, the
  probability computed from the lattice model
has a clear fat tail towards lower numbers of waters in the probe
volume.  This fat tail is the hallmark of a soft vapor-liquid interface, in this
case a soft interface next to
the hydrophobic solute.\cite{MittalHummer2008,GodawatJamadagniGarde2009,SarupriaGarde2009,MittalHummer2010}  At higher attractive
interactions, this fat tail is correspondingly depressed,
but not entirely suppressed.  Accordingly, in
Ref.~\onlinecite{PatelVarillyChandler2010}, the fat tail is only fully
suppressed when $\eta$ exceeds~$3.0$.

Figure~\ref{fig:PvNNextToPlate} also evidences some of the limitations
of the present model.  The probe volume being less than one lattice
cell thick, large lattice artifacts are inevitable.  Moreover, since
$P_V(N)$ distributions are much more detailed probes of solvent structure than
solvation free energies, we expect more room for disagreement with
simulation.  Nevertheless,
we emphasize that, by construction, no implicit solvation models can
capture the above effects on solvent structure, which underlie the
pathways of hydrophobic assembly.  Other coarse-grained solvation
models (for example, see Ref.~\onlinecite{SetnyZacharias2010}), on the
other hand, \emph{can} probe rare solvent fluctuations, and it would
be useful to evaluate their accuracy in this respect as compared to
explicit-water models and the present lattice model.

\subsection{Confinement}

\begin{figure}
\includegraphics{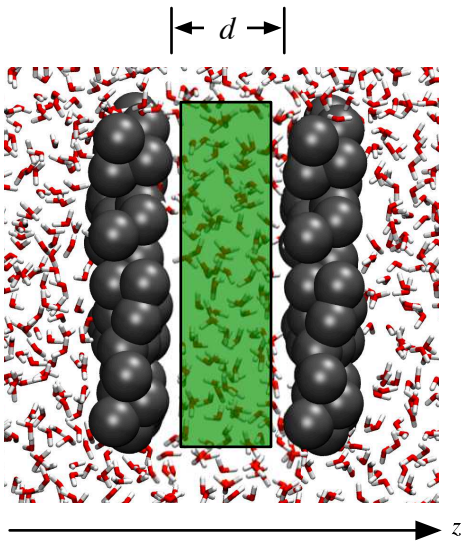}
\caption{Setup for examining water fluctuations under confinement
  (here, $d = 8\,$\AA).
  The model hydrophobic plates~\cite{PatelVarillyChandler2010} (grey
  particles) are placed~$d\,$\AA\ apart:  taking into account the van der Waals
  radii of about~$2\,$\AA\ of the plates' oily particles and the
  $3\,$\AA\ thickness of each plate, the
  center of the first plate is placed at $z=0$, the center of the
  second plate is placed at $z = d + 7\,$\AA.  The van der Waals
  radius of water (red and white sticks) being about~$1.5\,$\AA, the
  $24\times24\times(d-3)\,$\AA$^3$ probe volume (green) extends from
  $z=5\,\text{\AA}$~to~$z=d+ 2\,\text{\AA}$.  The plates are not
  perfectly flat, so some waters fit between the plates and the probe volume.\label{fig:TwoPlateSetup}}
\end{figure}
\begin{figure}
\includegraphics{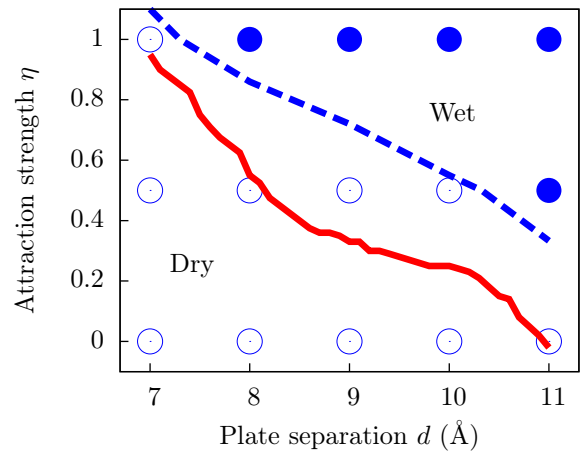}
\caption{\label{fig:WetDryPhaseDiagram}Phase diagram for the interplate region of the system
  depicted in Figure~\ref{fig:TwoPlateSetup}.  For the explicit SPC/E
  water simulations (blue), each symbol corresponds to
  an individual $P_V(N)$ distribution that we have calculated (filled:
  wet state stable; open: dry state stable).  The phase boundary (blue
  dashes) is estimated from a linear interpolation of the
  relative stability of the wet and dry states.  The relative
  stability is determined from the relative depths of the basins
  in~$-\ln P_V(N)$  The phase
  boundary for the present model (red solid line) was estimated from a dense sampling of $P_V(N)$ distributions, and is accurate to
  $\pm0.1\,$\AA~in~$d$ and $\pm 0.1$~in~$\eta$.}
\end{figure}

To examine confinement in detail, we place two of the model
hydrophobic plates at a distance~$d$ from each other, as
shown in Figure~\ref{fig:TwoPlateSetup}, and calculate the water
number distribution in a $24\times24\times (d-3)\,$\AA\ probe volume
between them as a function of interplate separation~$d$ and attraction
strength~$\eta$.  Figure~\ref{fig:WetDryPhaseDiagram} summarizes
the results in the form of a phase diagram.  At small separations and
low attractive strengths, the dry state (low~$N$) is most stable,
whereas high attractive strengths and large separations stabilize the wet
state (high~$N$).  Generically, the hydrophobic association of two
such plates proceeds through a dewetting transition in the inter-plate
volume.\cite{WallqvistBerne1995,LumLuzar1997,BolhuisChandler2000,MelittinNature2005}

\begin{figure}
\includegraphics{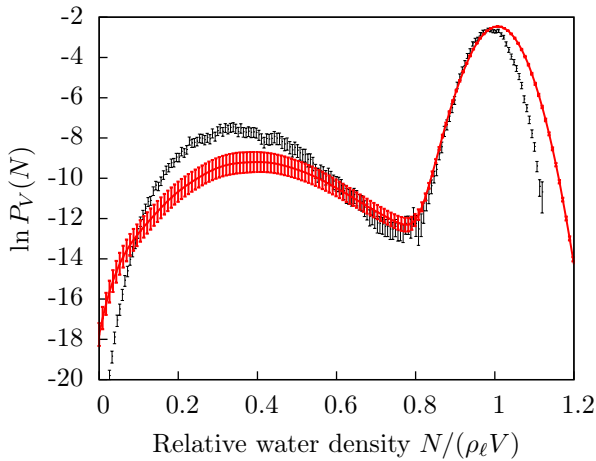}
\caption{\label{fig:PvNTwoPlateExample}Water density distribution of
  confined water~$1\,$\AA\ from coexistence.  These distributions are for
  the system depicted in Figure~\ref{fig:TwoPlateSetup} when
  $\eta=0.5$.  Coexistence lines are shown in Figure~\ref{fig:WetDryPhaseDiagram}.  The explicit water simulation data (black)
  corresponds to $d=11\,$\AA, while the coarse-grained model (red)
  results correspond to $d=9\,$\AA.  The remaining
$P_V(N)$ distributions are included in the
Supplementary Data\cite{SuppRef}.}
\end{figure}

The general, though not quantitative, agreement between the
coarse-grained model and
the SPC/E data is very encouraging: bistability is observed in the $P_V(N)$ distributions
in both cases, with the barriers at the nearly equal values of~$N$, and
with barrier heights that track the SPC/E barrier heights.  The phase
boundary in Figure~\ref{fig:WetDryPhaseDiagram} closely tracks the phase
boundary observed in explicit water, with a shift of less
than~$2\,$\AA\ for all~$\eta$.  Moreover, as shown in
Figure~\ref{fig:PvNTwoPlateExample}, once the general shift in the phase
boundaries is accounted for, the $P_V(N)$ distributions for systems
near that boundary obtained by
the coarse-grained model and the SPC/E simulations agree reasonably well.  Hence, the present model is better
suited than implicit solvation models for studies of nanoscale
self-assembly or protein-protein interactions driven by the
hydrophobic effect.  A recently developed coarse-grained
model of water (mW water) has been used to extensively probe these rare
fluctuations, and their predictions also display the characteristic
bistability of dewetting transitions that we observe.\cite{MolineroMoore2009,XuMolinero2010}

\section{Discussion}
\label{sec:discussion}

We have presented a coarse-grained model of solvation thermodynamics
that correctly reproduces the length-scale dependence of solvation
free energies, and, moreover, correctly captures the behavior of the
slow and rare solvent fluctuations that are pivotal in pathways to
hydrophobic assembly.  Our model is applicable to generic solute
shapes, and addresses the effects of solute-solvent attractive interactions.

While our model successfully describes various aspects of the
hydrophobic effect, several technical challenges must be addressed
before it can be applied in biological settings.  Most notably, electrostatic forces are missing from
our model.  As a first approximation, the GB treatment may well be
sufficiently accurate, as long as the low-permittivity cavity includes both
the solute's excluded volume and the regions where $n(\vr)$ is
zero.  It may also be possible to implement electrostatics in terms of a dipole density field coupled to the water density field.  The statistics of the dipole field are known to be Gaussian\cite{SongChandlerMarcus1996,SongChandler1998} so that their contribution to~$\Heff[\{n_i\}]$ may be computed analytically.

A second notable technical hurdle is to find efficient algorithms for
calculating the gradient of $\Heff[\{n_i\}]$ with respect to the position of
the solute's atomic centers, necessary for implementing realistic
solute dynamics, such as Brownian dynamics.  In the context of solvent lattice models, the
problem is tractable for spherical solutes with limited
overlap,\cite{tenWoldeChandler2002,WillardChandler2008} but the
implementation of a solution for generic solutes is more challenging.

Finally, as with implicit solvation models, our own model does not
attempt to capture solvent \emph{dynamics}.  For
thermodynamically-driven processes, almost any reasonable dynamics may
suffice when estimating the kinetic prefactor of rate constants of
interest.  Indeed, in a previous lattice model,\cite{tenWoldeChandler2002} the solvent dynamics is approximated by Glauber dynamics, the solute's by
Langevin dynamics, and the relative rates at which the two dynamics
are advanced are calibrated through physically reasonable arguments.
However, it is known, as evidenced in the form of the Oseen tensor,
that hydrodynamic interactions can be long-ranged\cite{DoiEdwards1988} and can influence timescales of molecular processes by one or more orders of magnitude.\cite{KikuchiEtAl2002}  This observation may prove important in nanoscale assembly processes that are kinetically driven, rather than thermodynamically driven.\cite{WhitelamEtAl2009}  Approaches to implementing coarse-grained dynamics in a lattice setting include multiparticle collision dynamics,\cite{MalevanetsKapral1998} fluctuating hydrodynamics\cite{Voulgarakis2009} and lattice Boltzmann methods,\cite{LatticeBoltzmannReview1998} among others.  We leave all dynamical considerations to future work.

\begin{acknowledgments}
  NIH Grant No.~R01-GM078102-04 supported P.V.~in the later
  stages of this work, A.P.~throughout and D.C.~in the early stages.  In the early stages, P.V.~was supported by a Berkeley
  Fellowship.  In the later
  stages, D.C.~was supported by the Director, Office of Science, Office of Basic Energy Sciences, Materials Sciences and Engineering Division and Chemical Sciences, Geosciences, and Biosciences Division of the U.S. Department of Energy under Contract No. DE-AC02-05CH11231.
  John Chodera and Michael Shirts helped in understanding and
  implementing MBAR.  We thank David Limmer, Ulf Pedersen and Thomas
  Speck for a critical reading of the manuscript.
\end{acknowledgments}

\appendix

\section{Derivation of the Model}
\label{app:derivation}

\subsection{Continuum formulation}
\label{app:derivation:continuum}

In this appendix, we derive Equation~\eqref{eqn:model} starting from
the microscopic ideas of LCW theory embodied in
Equation~\eqref{eqn:HMicro}.  The forms of $H_{\text{small}}[\delta\rho(\vr);n(\vr)]$~and~$H_{\text{large}}[n(\vr)]$ are given in
Equations \eqref{eqn:Hsmall}~and~\eqref{eqn:HLContinuum}, and are
discussed in the main text.  Here, we begin by discussing the details of the
$H_{\text{int}}[n(\vr),\delta\rho(\vr)]$ term.  Next, we integrate out the field~$\delta\rho(\vr)$ to obtain the effective
Hamiltonian~$\Heff[n(\vr)]$.  In the following section, we discretize it.

Whenever $n(\vr)$ is non-uniform, solvent molecules experience an
effective attraction towards denser regions, or equivalently, an
effective repulsion from less dense regions.  As argued by Weeks and
coworkers,\cite{Weeks2002,LumChandlerWeeks1999,tenWoldeSunChandler2001}
this effect can be modeled as a coupling, $H_{\text{int}}[n(\vr),\delta\rho(\vr)]$, between an
external \emph{unbalancing} potential, $\phi(\vr)$, and solvent density.  In the absence of a solute, the
energetics of this effect is completely contained in
$H_{\text{large}}[n(\vr)]$, but the presence of a solute gives rise to
important corrections.  Formally, the coupling is given by
\begin{equation}
H_{\text{int}}[n(\vr),\delta\rho(\vr)] = \int_\vr \phi(\vr)
\delta\rho(\vr) + \Hnorm[n(\vr)],\label{eqn:HPhiContinuum}
\end{equation}
where
\begin{equation}
\phi(\vr) = -2 a \rho_\ell [ \overline{n(\vr)} - 1 ].
\label{eqn:phiContinuous}
\end{equation}
Here, $a$ determines the strength of the potential, and the
overbar operator smears $n(\vr)$ over the effective range of
solvent-solvent attractive interactions.  The potential is shifted so
that it is zero for the uniform liquid.  The term $\Hnorm[n(\vr)]$ is 
chosen so that, in the absence of a solute,
$\Heff[n(\vr)]$ is identical to $H_{\text{large}}[n(\vr)]$.

We now integrate out the field $\delta\rho(\vr)$ to obtain~$\Heff[n(\vr)]$.  For notational simplicity, we
suppress the dependence of~$\chi(\vr,\vr')$ on $n(\vr)$ manifest in
Equation~\eqref{eqn:chi}.  The total density $\rho_\ell n(\vr) +
\delta\rho(\vr)$ is constrained to be zero for all points
$\vr$~in~$v$, so the effective Hamiltonian is given by
\begin{multline}
\exp\{-\beta \Heff[n(\vr)]\} = \int \mathcal{D}\delta\rho(\vr)\\
\exp\{-\beta H[n(\vr),\delta\rho(\vr)]\} \prod_\vrin \delta( \rho_\ell
n(\vr) + \delta\rho(\vr) ).
\label{eqn:HeffDefn}
\end{multline}
A long but straightforward calculation\cite{Chandler1993} yields
\begin{widetext}
\begin{multline}
\Heff[n(\vr)] =
H_{\text{large}}[n(\vr)]
+ \kB T\ln\sqrt{\det 2\pi\chiin}
- \int_\vrin \phi(\vr) \rho_\ell n(\vr)\\
+\frac{\kB T}{2} \int_\vrin \int_\vrPin
   \Biggl[ \rho_\ell n(\vr) - \int_\vrPPout \chi(\vr,\vr'')
     \beta\phi(\vr'') \Biggr]
   \chiin^{-1}(\vr,\vr')
   \Biggl[ \rho_\ell n(\vr') - \int_\vrPPPout \chi(\vr',\vr''')
     \beta\phi(\vr''') \Biggr]\\
- \frac{\kB T}{2} \int_{\vrout} \int_{\vrPout}
  \beta\phi(\vr)\chi(\vr,\vr')\beta\phi(\vr')
+ \Hnorm[n(\vr)].
\label{eqn:HeffSoluteOpaque}
\end{multline}
\end{widetext}
Here, $\chiin(\vr,\vr')$ is the restriction of $\chi(\vr,\vr')$ to the volume~$v$.  As such, $\chiin^{-1}(\vr,\vr')$ satisfies
\begin{equation}
\int_\vrPin \chiin^{-1}(\vr,\vr') \chi(\vr',\vr'') = \delta( \vr -
\vr''),\qquad \vr,\vr''\in v.
\label{eqn:chiin}
\end{equation}
To make $\Heff[n(\vr)]$ equal to $H_{\text{large}}[n(\vr)]$ in the absence of a
solute,
\begin{equation}
\Hnorm[n(\vr)] = \frac{\kB T}{2} \int_\vr \int_{\vr'}
\beta\phi(\vr)\chi(\vr,\vr')\beta\phi(\vr').
\label{eqn:Hnorm}
\end{equation}

It is useful to recast Equation~\eqref{eqn:HeffSoluteOpaque} into a
form where the physical significance of each term is manifest.  To do
so, we first note how the constraint of zero
solvent density inside~$v$ modifies the solvent density and its
fluctuation spectrum outside of~$v$.   As described in Ref.~\onlinecite{Chandler1993}, the average of $\delta\rho(\vr)\delta\rho(\vr')$ in the
presence of the constraint, denoted by~$\chim(\vr,\vr')$, is given by
\begin{multline}
\chim(\vr,\vr') = \chi(\vr,\vr')\\
 - \int_\vrPPin \int_\vrPPPin \chi(\vr,\vr'') \chiin^{-1}(\vr'',\vr''') \chi(\vr''',\vr').
\end{multline}
From Equation~\eqref{eqn:chiin}, it follows that $\chim(\vr,\vr')$ is zero
whenever $\vr$~or~$\vr'$ are in~$v$, as required by the solvent
exclusion constraint.  To describe the constraint's effect on the
average density, we introduce an auxiliary field~$c(\vr)$ that satisfies
\begin{subequations}
\begin{align}
\int_\vrPin \chi(\vr,\vr') c(\vr') &= \rho_\ell n(\vr),&\vr\in v,\label{eqn:cSimpleAC}\\
c(\vr) &= 0,&\vr\in\barvex.
\end{align}
\label{eqn:c}
\end{subequations}
In terms of $c(\vr)$~and~$\chim$, the average density in the presence of the solute is given by
\begin{multline}
\avg{\rho(\vr)} = \rho_\ell n(\vr) \\
-\int_{\vr'\in\vex} \chi(\vr,\vr') c(\vr') - \int_{\vr'\in\barvex} \chim(\vr,\vr') \beta\phi(\vr').
\label{eqn:avgRhoSimple}
\end{multline}

Equation~\eqref{eqn:HeffSoluteOpaque} can now be written much more simply as follows.
\begin{multline}
\Heff[n(\vr)] = H_{\text{large}}[n(\vr)]
 - \int_\vrin \phi(\vr) \rho_\ell n(\vr)\\
+ \kB T\ln\sqrt{\det 2\pi\chiin}
+\frac{\kB T}{2} \int_\vrin \rho_\ell n(\vr) c(\vr)\\
- \int_\vrout \int_\vrPin \phi(\vr) \chi(\vr,\vr') c(\vr')\\
+\frac{\kB T}{2} \int_\vr \int_{\vr'} \beta\phi(\vr)(\chi - \chim)(\vr,\vr')\beta\phi(\vr').
\label{eqn:HeffFullLLCW}
\end{multline}
For the geometries we considered, the sum of the last two terms of this equation
is, on average, opposite in sign but nearly proportional to the much
simpler remaining term involving~$\phi(\vr)$
(see Section~\ref{sec:HintRenorm}).  Physically, these three terms
capture the energetic bonus of driving~$\delta\rho(\vr)$ to~$0$
inside $v$ where $\phi$ is positive, the energetic cost of the
consequent density enhancement just outside of~$v$, and the small
difference between (a) the entropic cost associated with $\phi$ modifying the
solvent density in the presence of a solute and (b) that same cost in
the absence of a solute.  
In typical configurations, the three terms are
roughly proportional to the subvolume of~$\vex$ where $n(\vr) = 1$, and
capture how solvation free energies are modified by the microscopic
curvature of~$\vex$.   We have found it accurate to model the effect of these three
terms using \emph{only} the second term of
Equation~\eqref{eqn:HeffFullLLCW}, whose strength is then renormalized by a
factor~$K$.  The resulting approximation for $H_{\text{int}}[n(\vr)]$ is
\begin{equation}
H_{\text{int}}[n(\vr)] \approx -K \int_{r\in v} \phi(\vr) \rho_\ell
n(\vr).
\label{eqn:HintContinuous}
\end{equation}

Finally, we introduce an important simplification in
$H_{\text{small}}[n(\vr)]$.  Instead of solving Equation~\eqref{eqn:cSimpleAC} to
obtain the value of the field~$c(\vr)$ in~$v$, we replace $c(\vr)$ there by its average value,
$c_1$, and obtain the much simpler relation
\begin{equation}
c_1 = \avg{N}_v / \sigma_v,
\label{eqn:cOneBasis}
\end{equation}
where
\begin{align}
\avg{N}_v &= \int_{\vr\in v} \rho_\ell n(\vr),\label{eqn:continuousAvgNv}\\
\sigma_v &= \int_{\vr\in v} \int_{\vr'\in v} \chi(\vr,\vr').\label{eqn:continuousSigmaV}
\end{align}
Equivalently, in Equation~\eqref{eqn:HeffDefn} we enforce the single constraint that $\int_{\vr\in v}
\rho_\ell n(\vr) + \delta\rho(\vr)$ be zero, instead of enforcing
the multitude of constraints that $\rho_\ell
n(\vr) + \delta\rho(\vr)$ be zero at every point $\vr$~in~$v$.  We have verified
that this approximation, dubbed the ``one-basis set approximation'' in
previous works,\cite{tenWoldeSunChandler2001,HuangChandler2002} does
not appreciably change the solvation free energies and $P_V(N)$
distributions that we have obtained.  Crucially, this approximation
replaces the large (though sparse) linear system of
Equation~\eqref{eqn:cSimpleAC} with the trivial relation of
Equation~\eqref{eqn:cOneBasis}, and is therefore very advantageous
computationally.  With it, the term $H_{\text{small}}[n(\vr)]$ is
given by
\begin{equation}
H_{\text{small}}[n(\vr)] = \kB T [ \avg{N}_v^2/2\sigma_v + C/2 ],
\label{eqn:HsmallContinuous}
\end{equation}

The normalization constant~$C$ is defined by Equation~\eqref{eqn:C}.
When the value of~$\avg{N}_v$ becomes small, the integral
defining~$\sigma_v$ is dominated by the $\delta$-function in
Equation~\eqref{eqn:chi0} and takes the value $\sigma_v \approx
\avg{N}_v$.  The value $\ln(2\pi\sigma_v)$ of $C$ that is applicable for
larger~$\avg{N}_v$ thus tends unphysically to negative infinity as
$\avg{N}_v$ tends to zero.  This deficiency arises from a breakdown
of Gaussian statistics for solvent number fluctuations in sub-Angstrom volumes.  Since solvent
molecules are discrete entities, these statistics are instead
Poissonian.  A small cavity~$v$ can contain
either one solvent molecule, with probability $\avg{N}_v$, or no
solvent molecules, with probability $1 - \avg{N}_v$.  The free-energy
cost of evacuating that cavity is thus $-\kB T \ln(1 - \avg{N}_v)
\approx \avg{N}_v \kB T$.  The definition of $C$ given in
Equation~\eqref{eqn:C} is a simple, continuous way of
capturing this difference in fluctuation statistics at tiny
length-scales.  The crossover occurs at $\avg{N}_v \approx (2\pi - 2)^{-1} \approx 0.23$.

\subsection{Lattice formulation}
\label{app:derivation:lattice}

Using Equation~\eqref{eqn:nofr}, we express $\Heff[n(\vr)]$ and its
component terms in terms of the lattice variables~$n_i$, so that
\begin{equation}
\Heff[\{n_i\}] = H_{\text{large}}[\{n_i\}] + H_{\text{int}}[\{n_i\}] + H_{\text{small}}[\{n_i\}].\label{eqn:HEffSimpleDC}
\end{equation}
The integrals that define each term are then approximated through lattice
sums, with continuous fields replaced by either their average values or
their integrals over each cell.

Equation~\eqref{eqn:HsmallContinuous} for $H_{\text{small}}[\delta\rho(\vr);n(\vr)]$ is the easiest to tackle.  We begin by discretizing
Equation~\eqref{eqn:chi}, which defines~$\chi(\vr,\vr')$, when the
domains of integration for $\vr$~and~$\vr'$ are $V$~and~$V'$,
respectively.  In terms of the matrix~$\chi_{ij}(V,V')$ defined by
Equation~\eqref{eqn:Aij}, our prescription yields
$$
\chi(\vr,\vr') \to n_i \chi_{ij}(V,V') n_j,\qquad \vr\in V,\,
\vr'\in V'.
$$
Equation~\eqref{eqn:discreteSigmaV} for $\sigma_v$ then follows immediately
from Equation~\eqref{eqn:continuousSigmaV}.
Equation~\eqref{eqn:discreteAvgNv} for~$\avg{N}_v$ reasonably
approximates the integral in Equation~\eqref{eqn:continuousAvgNv}.

To discretize Equation~\eqref{eqn:HintContinuous} for~$H_{\text{int}}[n(\vr)]$, we need to choose a
concrete implementation of the overbar operation that is used to define~$\phi(\vr)$.  Following
Ref.~\onlinecite{tenWoldeSunChandler2001}, we approximate it as a
weighted average involving the cell and its nearest
neighbors\footnote{In Ref.~\onlinecite{tenWoldeSunChandler2001}, the
  term proportional to~$n_i$ is omitted.  Since~$\phi(\vr)$ only acts
  on cells with $n_i = 1$, this omission is inconsequential, and shows
  up as an extra factor of~$2$ in their value of~$a$}, given by
$$
\overline{n(\vr)} \to \Bigl[\frac12 n_i + \frac1{12} \sum_{j\,\text{(nn$i$)}} n_j\Bigr].
$$
The average, $\phi_i$, of $\phi(\vr)$ over cell~$i$ follows
immediately from Equation~\eqref{eqn:phiContinuous}, and is given by Equation~\eqref{eqn:discretePhii}.
Following our prescription, Equation~\eqref{eqn:HintContinuous} is
then reasonably discretized as the lattice sum
$$
H_{\text{int}}[\{n_i\}] \approx K \sum_i \phi_i (-\rho_\ell n_i \vexi).
$$

Discretizing $H_{\text{large}}[n(\vr)]$ correctly is a surprisingly subtle challenge.
Previously,\cite{tenWoldeSunChandler2001,tenWoldeChandler2002,WillardChandler2008}
it has been approximated it by an Ising Hamiltonian with nearest-neighbor
coupling
$$
H_{\text{large}}[\{n_i\}] \stackrel{?}{\to} \gamma \lambda^2 \sum_{\langle i j \rangle} (n_i - n_j)^2 - \mu\rho_\ell\lambda^3\sum_i n_i.
$$
Unfortunately, the use of this Hamiltonian results in serious
artifacts.  Consider, for instance, the energetics of a convex vapor
bubble embedded in the liquid, as represented by the field~$\{n_i\}$.
Many configurations of the field that are physically distinct have
nonetheless equal projections onto the $xy$-, $yz$-~and~$xz$-planes,
so they will be given equal statistical weight by the Hamiltonian.
Hence, the use of this Hamiltonian results in an unphysical excess of
entropy, as shown in detail in Section~\ref{app:tWC02Compare}.
Moreover, the energetic cost of common configurations of the
field~$\{n_i\}$ is substantially overestimated.  The Ising Hamiltonian
assigns a large vapor bubble of radius~$R$ an interfacial energy of
about~$6\pi \gamma R^2$, not $4\pi \gamma R^2$.  Whereas using a
renormalized~$\gamma$ can alleviate this latter problem,\cite{MillerVandenEijndenChandler2007} the problem
of excess entropy is more fundamental.

\begin{figure}
\begin{center}
\includegraphics{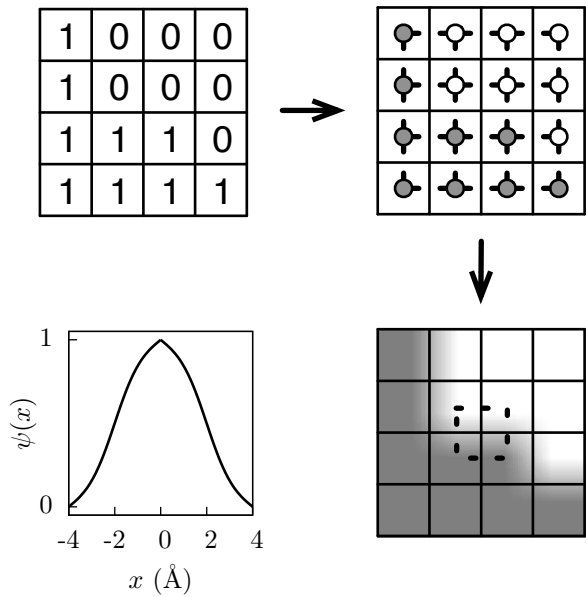}
\end{center}
\caption{Constructing $n(\vr)$~from~$\{n_i\}$.  The binary field
  specifies whether the density at the center of each lattice cell
  should be that of the liquid or that of the vapor.  Between cell
  centers, the density is interpolated using the basis
  function~$\psi(x)$ (whose form for water is shown in the lower left
  panel).  The dashed lines delineate the domain of integration of
  the local free energy $h_i$ given by Equation~\eqref{eqn:hi}.\label{fig:BuildingRhoS}}
\end{figure}

Motivated by the above deficiencies of the Ising Hamiltonian, we have
instead chosen to evaluate the Landau-Ginzburg integral in
Equation~\eqref{eqn:HLContinuum} numerically.  To proceed, we need to
construct the basis function $\Psi(\vr)$ used in
Equation~\eqref{eqn:nofr}.  Our choice, depicted in
Figure~\ref{fig:BuildingRhoS} for water, approximates the usual van der Waals construction%
\cite{[][. Chapter 3.]RowlinsonWidom1982} at a local level.
We first construct a 1D basis function~$\psi(x)$ satisfying
\begin{equation}
w'(\psi(x),0) - m \psi''(x) = 0,
\label{eqn:ODEpsi}
\end{equation}
with boundary conditions $\psi(0) = 1$~and~$\psi(\lambda) =
0$.  We then extend the range of~$\psi(x)$ and symmetrize it so that
\begin{equation}
\psi(x > \lambda) = 0,
\end{equation}
and
\begin{equation}
\psi(x < 0)=\psi(-x).
\end{equation}
Finally, the three-dimensional basis function $\Psi(\vr)$ is
constructed from the one-dimensional profiles $\psi(x)$ to give
$$
\Psi(x,y,z) = \psi(x) \psi(y) \psi(z).
$$
The field $n(\vr)$ constructed from Equation~\eqref{eqn:nofr} using this
basis function has many useful properties: the value of $n(\vr)$ at
the center of each cell~$i$ corresponds to the state encoded in~$n_i$;
the density interpolates smoothly between adjacent cells; and the
density profile of a configuration representing an axis-aligned wall,
where all $n_i$'s are $1$ on one side of a plane and $0$ on the
other, nearly reproduces the interface profile given by the van der
Waals construction.

For water, we use the function $w(n,\mu)$ given in
Equation~\eqref{eqn:omega}.  This choice results in both sides of Equation~\eqref{eqn:ODEpsi} being proportional to~$m$, so the
function $\psi(x)$ is independent of~$m$.  In the free van der Waals
theory, where the boundary conditions on Equation~\eqref{eqn:ODEpsi}
are $\psi(-\infty) = 1$~and~$\psi(+\infty) = 0$, the density profile~$\psi_0(z)$ that results is
$$
\psi_0(z) = [ 1 + \tanh(z/d) ] / 2,
$$
which accurately describes the average density profile of an SPC/E
water slab at ambient conditions.  The thickness parameter~$d$ can thus be
determined from simulation.  A complication due to capillary waves is
that $d$ grows logarithmically with simulation box size,\cite{Weeks1977,MittalHummer2008} so different authors quote different values of $d$:
$1.27\,$\AA\ for a $19\times19\,$\AA${}^2$ interface in
Ref.~\onlinecite{HuangChandler2002} and $1.54\,$\AA\ for a
$30\times30\,$\AA${}^2$ interface in
Ref.~\onlinecite{VegaDeMiguel2007}.  We choose the smaller value
because the instantaneous configuration of $n(\vr)$ should be blurred
only by small-scale fluctuations, not by large-scale capillary waves,
which correspond instead to different conformations of $n(\vr)$. 
The profile shown in Figure~\ref{fig:BuildingRhoS} corresponds to the solution of
Equation~\eqref{eqn:ODEpsi} when the more restrictive boundary
conditions described above are imposed, with $\lambda = 4\,$\AA\ and $d=1.27\,$\AA.

With concrete choices of $\Psi(\vr)$, $w(n,0)$, $m$~and~$\lambda$,
the integrals $h_i$ defined by Equation~\eqref{eqn:hi} can be
evaluated.  We discuss the choice of~$m$ below.  As outlined in the
main text, the value of $h_i$ depends only on the values of $n_j$ for
the $8$~cells~$j$ that share one of the corners of cell~$i$.  Out of
the 256 possible configurations of $\{n_j\}$, only $14$ are unique
when one accounts for reflection, rotation and inversion symmetry.
Thus, only $14$ distinct integrals need to be evaluated numerically.
This decomposition bears a strong
resemblance to the marching cubes algorithm\cite{LorensenCline1987}
that reconstructs interfaces in volumetric data, and is widely used in
computerized tomography.

In principle, the value of $m$ is related to the surface tension by the relation\cite{RowlinsonWidom1982}
\begin{equation}
\gamma = \int_0^\lambda \Bigl[ w(\psi(x),0) + m \psi'(x)^2 / 2 \Bigr]
\dee x.
\label{eqn:gammaFromM}
\end{equation}
On a lattice, as exemplified above by the Ising Hamiltonian, this choice results in perfect interfacial energies for
flat axis-aligned interfaces at the expense of more common curved
interfaces.  Thus, we instead choose~$m$ self-consistenly such that $\psi(x)$ satisfies
Equation~\eqref{eqn:ODEpsi} and the calculated interfacial energy of some
reference geometry of surface area~$A$ is $\gamma A$.
Equation~\eqref{eqn:gammaFromM} corresponds to a cubic reference
geometry.  Since curved surfaces are far more common than flat one in
realistic solutes, we instead use large spheres as our reference geometry.

For the specific form of $w(n)$ that we use for water, $h_i$ is
proportional to~$m$ and $\psi(x)$ is itself independent of~$m$.
The above self-consistent procedure can hence be implemented quite
simply.  We first calculate the $h_i$ quantities up to a factor
of~$m$, and then pick $m$ to obtain the correct interfacial energies.
The resulting values of $h_i$ are given in Table~\ref{tab:hi}.

\begin{table*}
\caption{Relative interfacial free energy~$h_i$ for each distinct
  neighboring cell configuration (diagrams
  after Ref.~\onlinecite{LorensenCline1987}).  Highlighted corners denote
  cells~$j$ with $n_j = 1$, whereas the others refer to cells with
  $n_j=0$; cell~$i$ is the lower-left corner in the back.   To aid the eye, a schematic of the implied liquid-vapor interface of each configuration is shown in orange.  The values of $h_i$ are inversion-symmetric: interchanging
  highlighted and unhighlighted corners yields the same interface and
  interfacial energy.  Also shown are the values of~$h_i$ that would reproduce the energetics of the standard Ising lattice gas, namely~$\gamma\lambda^2\sum_{\langle ij \rangle} (n_i - n_j)^2$.\label{tab:hi}}
\includegraphics{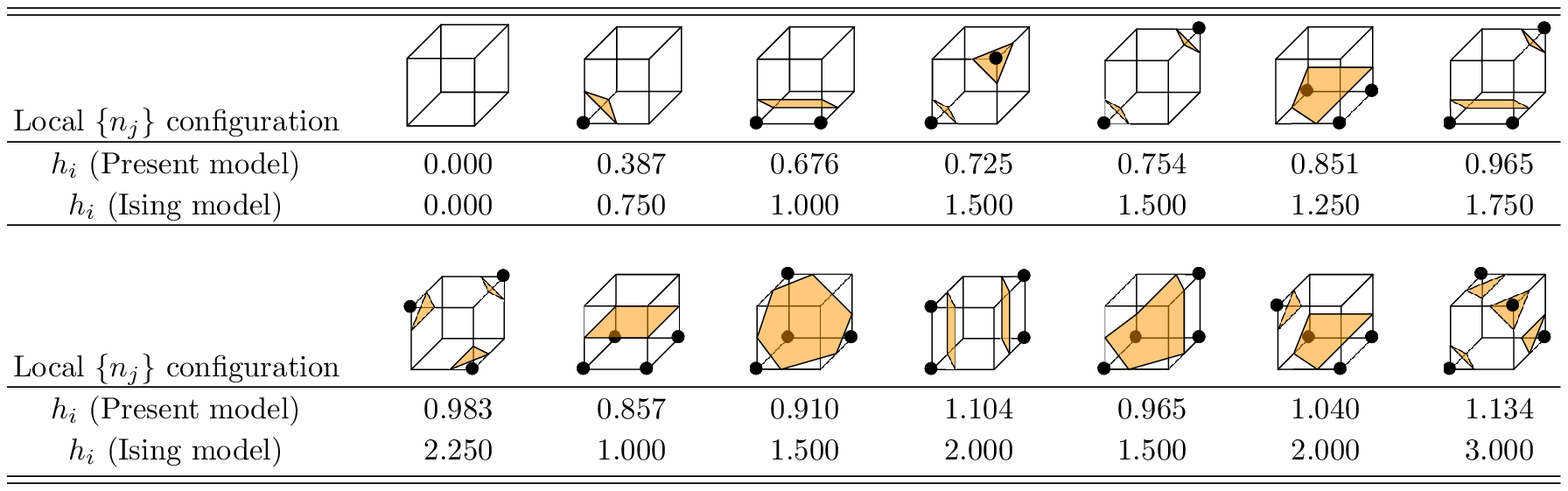}
\end{table*}

\subsection{Incorporating solute-solvent interactions}
\label{app:derivation:attractions}

A generic solute interacts with a solvent molecule through a
potential~$u(\vr)$.  This interaction is reflected in the microscopic Hamiltonian of
Equation~\eqref{eqn:HMicro} as an additional term~$H_u[n(\vr),\delta\rho(\vr)]$ given by
$$
H_u[n(\vr),\delta\rho(\vr)] = \int_\vr u(\vr) [\rho_\ell n(\vr) + \delta\rho(\vr)].
$$
Upon integrating out the density fluctuations, an additional term
$H_u[n(\vr)]$ appears in $\Heff[n(\vr)]$.  Physically, the total
solvent density responds linearly to the external field~$u(\vr)$
according to the density fluctuation spectrum is given by~$\chim(\vr,\vr')$, so that
\begin{multline}
\avg{\rho(\vr)} = \rho_\ell n(\vr) -\int_{\vr'} \chi(\vr,\vr') c(\vr')\\
 - \int_{\vr'} \chim(\vr,\vr') \beta[\phi(\vr') + u(\vr')].
\label{eqn:avgRhoWithUSimple}
\end{multline}
The resulting free energy change~$H_u[n(\vr)]$ arises from the direct
interaction of the solute and the solvent, and from the entropic cost
of modifying the mean solvent density around the solute.  It is given by
\begin{multline}
H_u[n(\vr)] = \int_\vr \dee\vr\, u(\vr) \avg{\rho(\vr)} \\
+ \frac{\kB T}{2} \int_\vr \int_{\vr'} \beta u(\vr) \chim(\vr,\vr') \beta u(\vr').
\label{eqn:HuContinuum}
\end{multline}
Note that the integrands are zero whenever $\vr$~or~$\vr'$ are inside
the solute.

To implement the previous equation on a lattice, we have found it useful to approximate~$\chim(\vr,\vr')$ by
\begin{equation}
\chim(\vr,\vr') \approx 
\begin{cases}
\chi_0(\vr - \vr'),&\vr,\vr' \in \barvex, n(\vr) = n(\vr') = 1,\\
0,&\text{otherwise}.
\end{cases}
\label{eqn:chimApprox}
\end{equation}
We also use the one-basis set approximation,
$c(\vr) \approx c_1$, given in Equation~\eqref{eqn:cOneBasis}.
Discretizing Equation~\eqref{eqn:HuContinuum} as in the previous
section then immediately yields Equation~\eqref{eqn:HuDiscrete}.

\section{Estimating~$\chi_{ij}(V,V')$}
\label{app:Aij}

An essential ingredient of the model we present is the
matrix~$\chi_{ij}(V,V')$, given by the integral in
Equation~\eqref{eqn:Aij}.  The terms involving the delta-functions of
Equation~\eqref{eqn:chi} are trivial.  Owing to the rapid oscillations
in $g(r)-1$, the remaining integrals are harder to estimate.  We
employ a two-step procedure to estimate these integrals efficiently.
We begin by subdiving the $\lambda = 4\,$\AA-resolution grid of cells
into a much finer grid of resolution~$\lambda_f = 1\,$\AA.  For
clarity, below we explicitly distinguish between cells in the
\emph{coarse} grid, indexed by the letters $i$~and~$j$, and cells in
the \emph{fine} grid, indexed by the letters $a$~and~$b$.  We evaluate
the integrals of the non-delta-function portion of $\chi_0$ on the
fine grid without otherwise restricting the arguments to particular
volumes $V$~and~$V'$, and denote the result by~$\chi_{ab}$.  Each fine cell is so small that the effect of a restriction on the integration domain can be estimated accurately with a simple interpolation formula.  We then use these interpolated values in the fine grid to build up the elements of $\chi_{ij}(V,V')$ over the coarse grid.

To evaluate $\chi_{ab}$, we use the Narten-Levy data for the structure factor~$S(k)$
of water.\cite{NartenLevy1971}  Since the $S(k)$ is unavailable for wave-numbers~$k$
higher than $16\,$\AA${}^{-1}$, we blur the domains of integration
over a range of about~$2\pi/16\,$\AA, which makes the values of the
integrals practically insensitive to this missing data.  Concretely,
we introduce a basis function~$\Phi$, given by
$$
\Phi(x,y,z) = \varphi(x) \varphi(y) \varphi(z),
$$
with
$$
\varphi(x) = \frac12 \left[ \tanh\frac{x-\lambda_f/2}{\Delta} - \tanh\frac{x+\lambda_f/2}{\Delta} \right].
$$
The function~$\varphi$ is unity around $x = 0$, and goes rapidly to
zero as $|x| \gtrsim \lambda_f/2$, with $\Delta$ controlling the range
of~$x$ over which this transition occurs.  We have found a value
of~$0.1\,$\AA\ for $\Delta$ to be adequate.  Using the notation $\vr_a$ to denote the center of fine cell~$a$, the value of $\chi_{ab}$ is given by
\begin{equation}
\chi_{ab} = \rho_\ell^2 \int_\vr \int_{\vr'}
\Phi(\vr - \vr_a) [ g(|\vr-\vr'|) - 1 ] \Phi(\vr' - \vr_b).
\label{eqn:chiab}
\end{equation}
The integral is best evaluated in Fourier space, where the term in
square brackets appears as the experimental~$S(k)$ profile.  We
overcome the convergence problems of a rapidly oscillating integrand
by using the Haselgrove-Conroy integration algorithm.\cite{Haselgrove1961,Conroy1967}  To properly account for $g(r)$
being exactly zero for $r \lesssim 2.35\,$\AA, we further set
$\chi_{ab}$ to exactly $-\rho_\ell^2$ if all points in~$a$ are within
$r_c = 2.35\,$\AA\ from all points in~$b$.  To limit the range of
$\chi_{ab}$, we also set it to zero if all points in~$a$ are more
than~$10\,$\AA\ from all point in~$b$.  The values of $\chi_{ab}$ need only be calculated once at each state point of water, and we have spent considerable effort in compiling them at ambient conditions.  Our results are included in the Supplementary Data\cite{SuppRef}.

For specific volumes $V$~and~$V'$, we estimate the value of $\chi_{ij}(V,V')$ as a weighted average
of the pertinent values of $\chi_{ab}$,
\begin{equation}
\chi_{ij}(V,V') \approx \rho_\ell (V \cap V') + \sum_{a \in i} \sum_{b \in j} 
(V_a /\lambda_f^3) \chi_{ab} (V'_b / \lambda_f^3),
\end{equation}
where $(V \cap V')$ is the volume of the overlap between $V$~and~$V'$.
This interpolation formula for
$\chi_{ij}$ is manifestly linear in its arguments, so that
$$
\chi_{ij}(V,V') + \chi_{ij}(V,V'') = \chi_{ij}(V,V' \cup V''),
$$
whenever $V'$~and~$V''$ do not overlap.  Most importantly, the interpolation procedure is simple, convenient, and correct for
the limiting cases of where all the values of $V_a$ are either
$0$~or~$\lambda_f^3$.

For comparison, we have also calculated values of $\chi_{ab}$ from an
explicit SPC/E water simulation in GROMACS at temperature $T = 298\,$K and
pressure $p=1\,$atm.  The values are also included in the
Supplementary Data\cite{SuppRef}.  For the
quantities we have studied in the main text, using
these values for $\chi_{ab}$ instead of those derived from the
Narten-Levy data yields nearly identical results.

\section{Fluctuation variance}
\label{app:chi}

The variance of the field $\delta\rho(\vr)$ given in Equation~\eqref{eqn:chi}
  is a simplification of the LCW interpolation formula,
$$
\chi_{\text{LCW}}(\vr,\vr') = \rho_\ell n(\vr) \delta(\vr - \vr') +
\rho_\ell^2 n(\vr) [g(|\vr-\vr'|) - 1] n(\vr'),
$$
to the case where $n(\vr)$ only takes the values $0$~or~$1$.  The
discrepancies arising from using
Equation~\eqref{eqn:chi} and more precise expressions for the variance
are mostly quantitative and limited to the vicinity of liquid-vapor
interfaces.\cite{Lum1998}

One possible improvement to Equation~\eqref{eqn:chi} is given in Ref.~\onlinecite{Chandler1993}:
\begin{multline}
\chi(\vr,\vr') = \chi_0(\vr,\vr') -\\
\int_{\vr''\in E} \dee\vr''\,
\int_{\vr'''\in E} \dee\vr'''\, \chi_0(\vr,\vr'')
\chi_E^{-1}(\vr'',\vr''') \chi_0(\vr''',\vr'),
\label{eqn:chiDC}
\end{multline}
where $E$ is the empty (i.e., gaseous) region of space where $n(\vr)$ is~$0$, and
$\chi_E^{-1}(\vr,\vr')$ satisfies
$$
\int_{\vr''\in E} \chi_E^{-1}(\vr,\vr'') \chi_0(\vr'',\vr') = \delta(\vr-\vr'),\quad \vr,\vr' \in E.
$$
Equations \eqref{eqn:chiDC}~and~\eqref{eqn:chi} are in qualitatively agreement:
both are zero when $\vr$ or $\vr'$ are in the gaseous region, and both reduce to
$\chi_0(\vr,\vr')$ well into the liquid phase.  The differences are, as
expected, concentrated near the boundaries of~$E$.  In this refined expression, the
integrand oscillates significantly within a lattice cell, so a lattice approximation to
Equation~\eqref{eqn:chiDC} proves unreliable.  Because using
Equation~\eqref{eqn:chi} gives accurate results for all quantities we
have examined, we regard the approximate Equation~\eqref{eqn:chi} to
be acceptable, and we have not pursued algorithms by which
Equation~\eqref{eqn:chiDC} can be accurately evaluated.

\section{How well is the effect of unbalanced forces captured by
  Equation~\eqref{eqn:HintContinuous}?}
\label{sec:HintRenorm}

Above, we replaced the three terms involving $\phi_i$ in
Equation~\eqref{eqn:HeffFullLLCW} by the simpler expression given in~\eqref{eqn:HintContinuous}.  We now justify this replacement.

Denote by $H_+[n(\vr)]$ the terms dropped from
Equation~\eqref{eqn:HeffFullLLCW}.  They are
\begin{multline}
H_+[n(\vr)] = - \int_\vrout \int_\vrPin \phi(\vr) \chi(\vr,\vr') c(\vr')\\
+\frac{\kB T}{2} \int_\vr \int_{\vr'} \beta\phi(\vr)(\chi - \chim)(\vr,\vr')\beta\phi(\vr').
\label{eqn:HPlus}
\end{multline}
Using the approximation for $\chim$ given in
Equation~\eqref{eqn:chimApprox} and the one-basis set approximation of
Equation~\eqref{eqn:cOneBasis}, we discretize these terms to obtain a
lattice version of $H_+[n(\vr)]$,
\begin{multline*}
H_+[\{n_i\}] = -\sum_{i,j} \phi_i n_i \chi_{ij}(\barvex,\vex) n_j \avg{N}_v/\sigma_v \\
+ \kB T \sum_{i,j} \beta\phi_i n_i [ \chi_{ij}(\vex,\vex) / 2 + \chi_{ij}(\barvex,\vex) ] n_j \beta \phi_j.
\end{multline*}
Because of the double sums in the formula, calculating $H_+[\{n_i\}]$ is by far the most computationally-demanding part of calculating $\Heff[\{n_i\}]$.  Since a single cell flip changes the value of~$\phi_i$ in up to~$7$ cells, calculating incremental changes to $H_+[\{n_i\}]$ is also much more expensive than calculating incremental changes to $H_u[\{n_i\}]$ (Equation~\eqref{eqn:HuDiscrete}), which has a similar structure.

\begin{figure}
\includegraphics{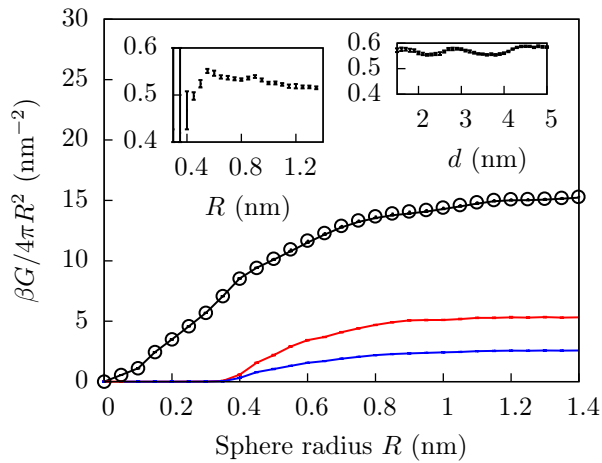}
\caption{Solvation free energies~$G$ of hard spheres
  as a function of sphere radius, where the term~$H_+[\{n_i\}]$
  (Equation~\eqref{eqn:HPlus}) has been included and the
  renormalization constant~$K$ has been set to~$1$ (solid black),
  compared to the simpler model in Equation~\eqref{eqn:model} (circles).  The averages of
  $-\langle H_{\text{int}}[\{n_i\}] \rangle$ (red) and $\langle H_+[\{n_i\}]\rangle$
  (black) are nearly proportional to each other.  Left Inset: implied
  renormalization constant~$K$, equal to $\avg{H_{\text{int}}[\{n_i\}] + H_+[\{n_i\}]}/\avg{H_{\text{int}}[\{n_i\}]}$.
  Note that both the numerator and denominator take on essentially zero value for $R \lesssim
  0.4\,$nm.  Right Inset: Implied value of~$K$ for hexagonal plate
  solute (Figure~\ref{fig:solvation_hexplates}) with $\eta=1.0$.  The implied value of $K$ is similar for different $\eta$.\label{fig:spheresFullLLCW}}
\end{figure}

Figure~\ref{fig:spheresFullLLCW} presents the solvation free energies
of hard spheres calculated when the $H_+[\{n_i\}]$ term is included and the
renormalization constant~$K$ is set to~$1$.  As can be seen, the term
corresponding to $H_{\text{int}}[\{n_i\}]$ has a much larger absolute value,
and in the region where their values are not negligible, the average
values of $H_{\text{int}}[\{n_i\}]$ and $H_+[\{n_i\}]$ are, as
claimed, essentially proportional.
The renormalization procedure we implement thus seems
justified, a conclusion borne out by the results in the text.
For completeness, we
have verified that the solvation free energies of the hexagonal plate
solute (Figure~\ref{fig:solvation_hexplates}) calculated when $H_+[\{n_i\}]$ is
included and $K$~is~$1$ are essentially identical to the ones
calculated using Equation~\eqref{eqn:model}.

\section{Comparison to the model of ten Wolde and Chandler}
\label{app:tWC02Compare}

Above, we argued that the Ising Hamiltonian estimate
for~$H_{\text{large}}[n(\vr)]$ overestimates the interfacial energy of a
sphere of radius~$R$ by a factor of $3/2$.  However, the lattice
version of LCW theory presented by ten Wolde and
Chandler\cite{tenWoldeChandler2002} uses precisely this Hamiltonian,
yet the solvation free energy of spheres seems to tend
to the correct value as $R$~grows.  Here we explain this apparent
paradox.

\begin{figure}
\includegraphics{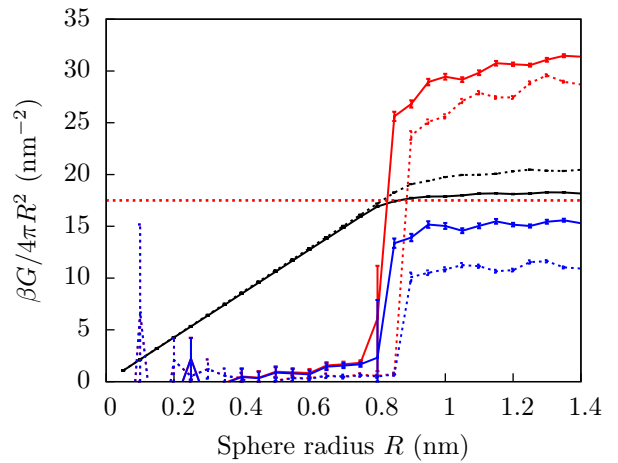}\\
\caption{Solvation free energies~$G$ of spheres in the model of
  Ref.~\onlinecite{tenWoldeChandler2002} (black), for cell
  sizes~$\lambda=2.1\,$\AA\ (solid) and $\lambda=2.3\,$\AA\ (dashes).  The use of the Ising Hamiltonian causes the
  average value of $H_{\text{large}}[\{n_i\}]$ (red) to significantly exceed
  the solvation free energy, but also leads to large excess entropies
  (blue, $TS = \avg{H} - G$).  At $\lambda = 2.1\,$\AA, but not at
  $\lambda = 2.3\,$\AA, a fortuitous
  cancellation leads to correct solvation free energies.\label{fig:tWC02}}
\end{figure}

Figure~\ref{fig:tWC02} shows the solvation free energies of spheres in
the model of Ref.~\onlinecite{tenWoldeChandler2002}, and shows how
they differ when the lattice cell size of $\lambda = 2.1\,$\AA\ is
changed to $\lambda = 2.3\,$\AA.  As claimed, 
$\avg{H_{\text{large}}[\{n_i\}]}$ is much larger than it should be, but
for $\lambda=2.1\,$\AA, the excess entropy resulting from the
unphysical degeneracies of the Ising Hamiltonian exactly cancels this excess energy.  This fortuitous cancellation does not occur for different cell sizes, and will not, in general, hold for solutes of different geometries.

%

\end{document}